# The collision between two ortho-Positronium (Ps) atoms: a four-body Coulomb problem

Hasi Ray[1,2,3]

[1] Study Center, S-1/407/6, B. P. Township, Kolkata 700094, India
[2] Department of Physics, New Alipore College, Kolkata 700053, India
[3] Science Department, National Institute of TTT and Research, Kolkata 700106, India

Email: hasi_ray@yahoo.com and hasi.ray1@gmail.com

**Abstract**: The elastic collision between two ortho-positronium (e.g. S=1) atoms is studied using an ab-initio static-exchange model (SEM) in the center of mass (CM) frame considering the system as a four-body Coulomb problem where all the Coulomb interaction terms in the direct and exchange channels are treated exactly. A coupled-channel methodology in momentum space is used to solve Lippman-Schwinger equation following the integral approach. A new SEM code is developed in which the Born-Oppenheimer (BO) scattering amplitude acts as input to derive the SEM amplitude adapting the partial wave analysis. The s-, p- and d- wave elastic phase shifts and the corresponding partial cross sections for the spin alignment S=0, i.e. singlet (+) and S=2, i.e. triplet (-) states are studied. An augmented-Born approximation is used to include the contribution of higher partial waves more accurately to determine the total/integrated elastic cross section ($\sigma$), the quenching cross section ($\sigma_q$) and ortho to para conversion ratio ($\sigma/\sigma_q$). The effective range theory is used to determine the scattering lengths and effective ranges in the s-wave elastic scattering. The theory includes the non-adiabatic short-range effects due to exchange.

1. **Introduction**

   The scattering of two positronium (Ps) atoms is a problem of fundamental physical importance [1-17] and of relevance to the formation of the $Ps_2$ molecule [1-9] as well as current endeavours towards the production of Ps Bose-Einstien condensation (BEC) [1,10-11]. It is again useful for its applications in technology [18-19]. The Ps-Ps scattering problem is very difficult to treat theoretically as it is a 4-body Coulomb problem. At lower energies the particles are highly correlated. The proper symmetry of a two-Ps system is a quite complicated group theoretical problem [4-5]. However, in scattering theory when two ortho-Ps ( e.g. S=1) atoms collide, the total aligned spin of the system can be S=2 when both the atoms are aligned in the same direction (e.g. polarised beam) and S=0, if they are oppositely aligned. An ab-initio static-exchange model (SEM) [20-25] is used to study the collision between two ortho-Ps atoms following a coupled-channel methodology in the momentum space [20-26]. An integral approach to solve the non-relativistic Schrodinger equation e.g. the Lippman-Schwinger type equations, is used to get the coupled integral equations for different channels. The system is treated in the center of mass (CM) frame considering it as a four-center Coulomb problem [25,27]. All the Coulomb-interaction terms in both the direct and exchange/ rearrangement channels are treated exactly. In Ps-Ps system, all the four charges are of equal masses; in addition the two electrons and two positrons are indistinguishable. So the exchange between two electrons and between two positrons are equally probable. As the system is treated in the CM frame, the exchange between only two positrons and exchange between only two electrons are equivalent. Again if both the exchanges happen simultaneously, then the system is equivalent to the direct channel. So effectively there are two direct channels and two exchange channels. So the exchange between the electrons only, is sufficient to describe the collision process. Accordingly the spin aligned S=2 state as triplet (-) and spin aligned S=0 state as singlet (+) are described. The plus and minus signs in the bracket indicate that the space-parts of the system wavefunction are symmetric and antisymmetric respectively. Accordingly the total cross section could be defined taking one-fourth contribution from singlet and three-fourth contribution from triplet channels. However a multiplicative factor greater than one may be required due to larger reaction rate for both the exchanges between the electrons and between the positrons. The theory by its nature, includes the non-adiabatic short-range effects due to exchange between electrons as well as between positrons.

   In a system of four charged particles there are six Coulomb interaction terms. In the present system, two positron-electron interaction terms are used to define two Ps-atomic wavefunctions. So effectively there are four Coulomb interaction terms between the atoms. The simplest first-Born matrix elements are all nine-dimensional integrals. The analytical part of the matrix elements are relatively easier to compute in direct channel due to symmetry of system wavefunctions in the initial and final states; it is possible to get the closed form. But in the rearrangement channel when the electrons exchange their positions the initial and final state wavefunctions are not symmetric. It is extremely difficult to compute the nine-dimensional space-integration to include the effect of exchange/ antisymmetry of the system electrons exactly. Using the Fourier transform and Bethe integral formulas one can reduce the dimension of these matrix elements into tractable two-dimensional (X-Y) integrals with limits zero (0) to one (1). The positron-positron and two electron-positron interaction terms are relatively less difficult to compute due to presence of two coupled terms. But the electron-electron correlation matrix element contains three coupled terms and was extremely difficult to transform into a tractable two-dimensional form. It was less difficult in Ps and H system [20]; due to the very light mass of Ps it was a good approximation to consider the H-nucleus as the origin of center of mass and it was treated as a three-center problem. By using a simple substitution of variables and accordingly changing the space of integration, it has been possible to transform the electron-electron exchange integral in the four-center problem into a nine-dimensional integral having only two coupled terms that can be reduced into a tractable similar two-dimensional form following exactly the same procedure as is used in other correlation terms.

The present paper deals the elastic collision between two ortho-Ps atoms when both are in ground (1s) states. The eigen-state expansion method to define the system wavefunction and an approach like the Hartree-Fock variational method to project out different channels are used to get the coupled integro-differential equations. One can use iterative method to find the unknown coefficients used in eigen-state expansion with partial wave analysis. However the number of coupled equations is restricted here by the number of bound states [26] taken into account. This is known as differential approach. In integral approach one uses the help of Lippman-Schwinger equation to get the coupled integral equations [26]. Either the configuration space [27] or the momentum space [20-26] can be used to form the equations. However in the momentum space formalism [20] the convergence problem is much easier to overcome than the coordinate space formalism. When the exchange amplitude is combined (adding or subtracting) with the direct first-Born amplitude, it is called the Born-Oppenhimer (BO) amplitude following the nomenclature of Ray and Ghosh [20] and accordingly the singlet (+) and triplet (-) channels are defined. The BO ($\pm$) amplitudes are used as input in the coupled-channel methodology to derive the SEM amplitudes for the singlet (+) and triplet (-) channels respectively. A partial wave analysis and angular momentum algebra are used to reduce the three-dimensional coupled integral equation into the one-dimensional form. The partial wave contributions: L=0 is defined as the s-wave, L=1 as the p-wave and so on. The s-, p-, d- wave elastic phase shifts, the corresponding s-, p-, d- wave elastic cross sections, the integrated/ total elastic cross section, the quenching cross section and ortho to para conversion ratio are evaluated. To calculate the integrated/ total elastic cross section ($\sigma$) more accurately to include the contribution of higher partial waves, an augmented-Born approximation [20] is used. The effective range theory [28] is applied to derive the scattering length ($a$) and the effective range ($r_0$) utilizing the variation of the s-wave elastic phase-shift ($\delta_0$) with the incident energy. To get the scattering cross section one has to integrate the square of the amplitude over the scattering angle [$\Omega(\theta,\varphi)$]. Accordingly an efficient computer-code is developed to compute the SEM amplitude utilizing the Born-Oppenheimer (BO) amplitude as input. It could be useful as the starting code to solve many challenging problems involving two-atomic collision.

2. **Theory**

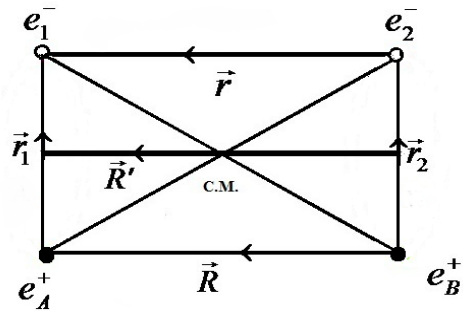

**Figure 1. The picture of Ps and Ps system.**

The description of the present Ps and Ps system is made in **Figure 1**. The vector $\vec{r}_1$ represents the relative coordinate between the positron $e_A^+$ and the electron $e_1^-$ in first Ps. Similarly vector $\vec{r}_2$ represents the relative coordinate between the positron $e_B^+$ and the electron $e_2^-$ in second Ps. $\vec{R}$ and $\vec{r}$ represent the relative coordinates between two positrons $e_A^+$ and $e_B^+$ and between two electrons $e_1^-$ and $e_2^-$, respectively. The initial and final state wavefunctions of the system are defined as

$$\psi_i(\vec{r}_1,\vec{r}_2,\vec{R}) = e^{i\vec{k}_i \cdot \vec{R}} \eta_{1s}^A(\vec{r}_1)\eta_{1s}^B(\vec{r}_2) \quad (2.1a),$$

$$\psi_f(\vec{r}_1,\vec{r}_2,\vec{R}) = (1 \pm P_{12})e^{i\vec{k}_f \cdot \vec{R}_f} \eta_{1s}^A(\vec{r}_1)\eta_{1s}^B(\vec{r}_2) \quad (2.1b),$$

so that $\vec{k}_i$ and $\vec{k}_f$ represent the initial and final momenta of the system (e.g. momenta of the projectile Ps if the target Ps is taken at rest); $\eta_{1s}^A(\vec{r}_1)$ and $\eta_{1s}^B(\vec{r}_2)$ represent the ground state wavefunctions of two Ps atoms; $P_{12}$ is the exchange or antisymmetry operator. $\vec{R}'$ and $\vec{R}_f$ represent the CM-coordinates of the system in the direct and exchange channels respectively and are defined as

$$\vec{R}' = \vec{R} + \frac{m_e}{m_A + m_e}\vec{r}_1 - \frac{m_e}{m_B + m_e}\vec{r}_2 \quad (2.2a),$$

$$\vec{R}_f = \vec{R} + \frac{m_e}{m_A + m_e}(\vec{r}_2 - \vec{R}) - \frac{m_e}{m_B + m_e}(\vec{r}_1 + \vec{R}) \quad (2.2b),$$

when $m_A$, $m_B$ are the masses of two positrons and $m_e$ is the electron mass.

In elastic scattering $|\vec{k}_i| = |\vec{k}_f|$; so only the direction of the final momentum $\vec{k}_f$ changes due to scattering. The Coulomb interaction between the atoms in the direct and exchange channels are expressed as

$$V_{Direct}(\vec{R},\vec{r}_1,\vec{r}_2) = \frac{e_A^+ e_B^+}{|\vec{R}|} - \frac{e_A^+ e_2^-}{|\vec{R} - \vec{r}_2|} - \frac{e_B^+ e_1^-}{|\vec{R} + \vec{r}_1|} + \frac{e_1^- e_2^-}{|\vec{R} + \vec{r}_1 - \vec{r}_2|} \quad (2.3a),$$

$$V_{Exchange}(\vec{R},\vec{r}_1,\vec{r}_2) = \frac{e_A^+ e_B^+}{|\vec{R}|} - \frac{e_A^+ e_1^-}{|\vec{r}_1|} - \frac{e_B^+ e_2^-}{|\vec{r}_2|} + \frac{e_1^- e_2^-}{|\vec{R} + \vec{r}_1 - \vec{r}_2|} \quad (2.3b),$$

respectively. Here the magnitudes of all the Coulomb terms in the numerators in equation (2.3a) and (2.3b) are equal to unity in atomic unit (a.u.). The formally exact Lippman-Schwinger type coupled integral equation for the scattering amplitude in momentum space [26] is given by :

$$f_{n'1s,n1s}^{\pm}(\vec{k}_f,\vec{k}_i) = B_{n'1s,n1s}^{\pm}(\vec{k}_f,\vec{k}_i) - \frac{1}{2\pi^2}\sum_{n''}\int d\vec{k}'' \frac{B_{n'1s,n''1s}^{\pm}(\vec{k}_f,\vec{k}'')f_{n''1s,n1s}^{\pm}(\vec{k}'',\vec{k}_i)}{\vec{k}_{n''1s}^2 - \vec{k}''^2 + i\varepsilon} \quad (2.4).$$

Here $B^{\pm}$ are the well known Born-Oppenheimer (BO) scattering amplitude [20-25] in the singlet (+) and triplet (-) channels respectively. Similarly $f^{\pm}$ indicate the unknown SEM scattering amplitudes for the singlet and triplet channels. The BO amplitude is defined as

$$B_{n'1s,n1s}^{\pm}(\vec{k}_f,\vec{k}_i) = -\frac{\mu}{2\pi}\int d\vec{R}d\vec{r}_1 d\vec{r}_2 \psi_f^*(\vec{R},\vec{r}_1,\vec{r}_2)V(\vec{R},\vec{r}_1,\vec{r}_2)\psi_i(\vec{R},\vec{r}_1,\vec{r}_2) \quad (2.5),$$

when $\mu$ is the reduced mass of the system. The four Coulomb interaction terms in equation (2.3a) and equation (2.3b): the first terms are the nucleus-nucleus (NN) interaction, the fourth terms are the electron-electron ($e_1 e_2$) interaction in both the channels; the second terms are the interaction between nucleus $e_A^+$ and electron $e_2^-$ ($Ae_2$) in direct channel and interaction between nucleus $e_A^+$ and electron $e_1^-$ ($Ae_1$) in exchange channel; similarly the third terms are the interaction between nucleus $e_B^+$ and electron $e_1^-$ ($Be_1$) in direct channel and the interaction between nucleus $e_B^+$ and electron $e_2^-$ ($Be_2$) in exchange channel.

The explicit form of the first term in the direct ($F_B^{NN}$) and exchange ($F_O^{NN}$) channels are:

$$F_B^{NN} = -\frac{\mu}{2\pi}\int d\vec{R}d\vec{r}_1 d\vec{r}_2 e^{-i\vec{k}_f\cdot\vec{R}}\phi_{1s}^{A*}(\vec{r}_1)\phi_{1s}^{B*}(\vec{r}_2)\frac{1}{R}e^{i\vec{k}_i\cdot\vec{R}}\phi_{1s}^A(\vec{r}_1)\phi_{1s}^B(\vec{r}_2) \quad (2.6a),$$

$$F_O^{NN} = -\frac{\mu}{2\pi}\int d\vec{R}d\vec{r}_1 d\vec{r}_2 e^{-i\vec{k}_f\cdot\vec{R}'}\phi_{1s}^{A*}(\vec{R}-\vec{r}_2)\phi_{1s}^{B*}(\vec{R}+\vec{r}_1)\frac{1}{R}e^{i\vec{k}_i\cdot\vec{R}}\phi_{1s}^A(\vec{r}_1)\phi_{1s}^B(\vec{r}_2) \quad (2.6b),$$

respectively. Similarly the fourth electron-electron correlation terms for the direct and exchange channels are,

$$F_B^{e_1 e_2} = -\frac{\mu}{2\pi}\int d\vec{R}d\vec{r}_1 d\vec{r}_2 e^{-i\vec{k}_f\cdot\vec{R}}\phi_{1s}^{A*}(\vec{r}_1)\phi_{1s}^{B*}(\vec{r}_2)\frac{1}{|\vec{R}+\vec{r}_1-\vec{r}_2|}e^{i\vec{k}_i\cdot\vec{R}}\phi_{1s}^A(\vec{r}_1)\phi_{1s}^B(\vec{r}_2) \quad (2.7a),$$

$$F_O^{e_1 e_2} = -\frac{\mu}{2\pi}\int d\vec{R}d\vec{r}_1 d\vec{r}_2 e^{-i\vec{k}_f\cdot\vec{R}'}\phi_{1s}^{A*}(\vec{R}-\vec{r}_2)\phi_{1s}^{B*}(\vec{R}+\vec{r}_1)\frac{1}{|\vec{R}+\vec{r}_1-\vec{r}_2|}e^{i\vec{k}_i\cdot\vec{R}}\phi_{1s}^A(\vec{r}_1)\phi_{1s}^B(\vec{r}_2) \quad (2.7b),$$

respectively.

The second and third terms in direct and exchange channels are:

$$F_B^{Ae_2} = -\frac{\mu}{2\pi}\int d\vec{R}d\vec{r}_1 d\vec{r}_2 e^{-i\vec{k}_f\cdot\vec{R}}\phi_{1s}^{A*}(\vec{r}_1)\phi_{1s}^{B*}(\vec{r}_2)\frac{(-1)}{|\vec{R}-\vec{r}_2|}e^{i\vec{k}_i\cdot\vec{R}}\phi_{1s}^A(\vec{r}_1)\phi_{1s}^B(\vec{r}_2) \quad (2.8a),$$

$$F_O^{Ae_1} = -\frac{\mu}{2\pi}\int d\vec{R}d\vec{r}_1 d\vec{r}_2 e^{-i\vec{k}_f\cdot\vec{R}'}\phi_{1s}^{A*}(\vec{R}-\vec{r}_2)\phi_{1s}^{B*}(\vec{R}+\vec{r}_1)\frac{(-1)}{|\vec{r}_1|}e^{i\vec{k}_i\cdot\vec{R}}\phi_{1s}^A(\vec{r}_1)\phi_{1s}^B(\vec{r}_2) \quad (2.8b),$$

$$F_B^{Be_1} = -\frac{\mu}{2\pi}\int d\vec{R}d\vec{r}_1 d\vec{r}_2 e^{-i\vec{k}_f\cdot\vec{R}}\phi_{1s}^{A*}(\vec{r}_1)\phi_{1s}^{B*}(\vec{r}_2)\frac{(-1)}{|\vec{R}+\vec{r}_1|}e^{i\vec{k}_i\cdot\vec{R}}\phi_{1s}^A(\vec{r}_1)\phi_{1s}^B(\vec{r}_2) \quad (2.9a),$$

$$F_O^{Be_2} = \frac{\mu}{2\pi}\int d\vec{R}d\vec{r}_1 d\vec{r}_2 e^{-i\vec{k}_f\cdot\vec{R}'}\phi_{1s}^{A*}(\vec{R}-\vec{r}_2)\phi_{1s}^{B*}(\vec{R}+\vec{r}_1)\frac{(-1)}{|\vec{r}_2|}e^{i\vec{k}_i\cdot\vec{R}}\phi_{1s}^A(\vec{r}_1)\phi_{1s}^B(\vec{r}_2) \quad (2.9b).$$

The most difficult term is the integral in Eqn.(2.7b) i.e. the electron-electron correlation with exchange; it contains three completely different coupled terms and two uncoupled terms in a nine-dimensional integral. The total elastic cross section ($\sigma$) is defined as

$$\frac{d\sigma}{d\Omega} = \frac{3}{4}|f^-|^2 + \frac{1}{4}|f^+|^2 \quad (2.10),$$

and the quenching cross section ($\sigma_q$) is defined as

$$\frac{d\sigma_q}{d\Omega} = \frac{1}{16}|f^- - f^+|^2 \quad (2.11).$$

The conversion ratio ($\sigma_q/\sigma$) is a parameter to quantify the fraction of total cross section transformed into para-Ps that could be measured. To calculate the integrated/ total elastic cross section more accurately an augmented-Born approximation is used. Here the contributions of higher partial waves are replaced by the equivalent first-Born term using the formulation

$$\sigma = \sum_l \sigma_l + \sigma^{BO} - \sum_l \sigma_l^{BO} \quad (2.10),$$

if $\sigma_l$ is the partial cross section using SEM; $\sigma^{BO}$ and $\sigma_l^{BO}$ are the total and partial BO cross sections.

### 3. Methodology

A highly efficient computer code is developed using the FORTRAN programming language and numerical analysis. The code calculates the elastic phase-shifts and corresponding cross sections for both the singlet (+) and the triplet (-) channels. Two sets of coupled one-dimensional integral equation one $f_L^+$ corresponding to symmetric space part and the other $f_L^-$ corresponding to antisymmetric space part are solved parallelly in the same code using the matrix inversion method for each partial wave (L). The pole term in the coupled integral equation (2.4) is evaluated using the formulation with delta function and principal value parts so that

$$\frac{1}{\vec{k}_n^2 - \vec{k}''^2 + i\varepsilon} = -i\pi\delta(\vec{k}_n^2 - \vec{k}''^2) + \frac{P}{\vec{k}_n^2 - \vec{k}''^2} \qquad (3.1).$$

The principal value integral from zero to infinity has been replaced by

$$\int_0^\infty dk'' = \int_0^{2k_n} dk'' + \int_{2k_n}^\infty dk'' \qquad (3.2).$$

Even number of Gaussian points in the interval $0 - 2k_n$ are used to avoid the singularity problem at $k'' = k_n$. The Gauss-Legendre quadratures are used to perform the two-dimensional integrations in $B^\pm$ and the $\theta$ integration numerically in the present code. All the integrations converged properly if the values of $k \leq 10$ a.u. The convergence is studied very carefully to integrate over the scattering angle ($\theta$) and the variables in the two-dimensional integrals (X and Y), varying the number of Gauss-Legendre quadratures. To study the correctness of the partial wave analysis method, the same partial wave analysis is applied to calculate the BO amplitude. It is verified that the BO amplitude obtained by partial wave analysis method and by the non-partial wave method are equal. More partial waves are required to get the converged data at higher energies. To include the contribution of the higher partial waves as accurately as possible in the integrated cross section, an augmented Born approximation [20-25] is used in the present code. To apply the augmented-Born method, we need to compare the SEM amplitude with the corresponding BO amplitude for each partial wave (L). As the value of L increases the difference between the two decreases gradually. When both are almost equal, only then it can replace the higher partial wave contribution of SEM amplitude by the equivalent BO amplitude. It is possible to get reliable data upto the value of $k \leq 10$ a.u. i.e. the energy E=1360 eV in Ps-Ps scattering using a coupled-channel methodology.

To determine the s-wave elastic scattering lengths ($a^+$ and $a^-$), the effective range theory that expresses s-wave elastic phase shift ($\delta_0$) as a function of scattering-length ($a$) and projectile energy ($\sim k^2$) so that

$$k \cot \delta_0 = -\frac{1}{a} + \frac{1}{2} r_0 k^2 + O(k^4) \qquad (3.3),$$

is useful, when $k$ is the magnitude of the incident momentum in atomic unit and $r_0$ is the effective range. The incident energy is related to $k$ by the relation $E(eV) = (27.21k^2/2\mu)$. The negative scattering length ($a\langle 0$), means no possibility of binding in the system. But the positive scattering-length ($a\rangle 0$) means that a binding is physically possible; so it indicates the possibility of Feshbach resonances and the BEC formation. A rapid change in phase-shift by $\pi$ radian in a very narrow energy interval is an indication of the presence of a Feshbach resonance [23-24, 28]; but the resonance should be reflected in the s-wave partial cross sections too. So in the present code, the partial cross sections are derived directly from the real and imaginary parts of the partial amplitudes avoiding the phase shifts. The presence of a Feshbach resonance indicates a binding in the system. To study the resonances, a very large number of mesh-points in a very narrow energy interval is necessary. So a good computation facility is required for the purpose.

### 4. Results and discussion

A new code is developed to study the elastic collision between two ortho-Ps atoms (when both are in ground states) using the ab-initio SEM theory. Here the CM frame is used to study the four body Coulomb problem of equal masses. In the new code, the BO amplitude acts as input to derive the unknown SEM amplitude. All the Coulomb interaction terms in both the direct and exchange channels to derive the BO amplitude are calculated exactly. The theory is useful to include the non-adiabatic short-range effects due to exchange. The detailed data obtained by using the present new code are being reported. The s-wave elastic phase shifts of both the total spin aligned states: S=0 for the singlet (+) and S=2 for the triplet (-) are presented in **Figure 2** and the corresponding s-wave partial cross sections are plotted in **Figure 3** against the incident momenta $k = 0.1$ to $k = 0.6$ a.u. at the energy region below the threshold. These data are compared with the data of Ivanov et al [14] in **Figure 2** and are almost similar to the data reported earlier [13] using similar type of calculation but approximating the electron-electron exchange term. In **Table 1a**, the s-, p- and d- wave elastic phase shifts and in **Table 1b**, the corresponding s-, p- and d- wave partial cross sections are presented following the present exact-exchange analysis. It should be noted that all the p-wave (odd parity) phase shifts are zero or $\pm \pi$ radian and all the p-wave cross sections are zero.

In the present new code when the most difficult electron-electron interaction matrix-element is approximated by the positron-positron interaction matrix-element, the s- (even parity) and d- (even parity) wave data are exactly the same with the exact-exchange analysis data of **Table 1a and 1b**, but all the p-wave (odd parity) data are very different and partial cross sections are not zero or close to zero. The present s-, p- and d- wave data using the approximate form for the electron-electron exchange term are presented in **Tables 2a** and **2b**. Comparing the phase shifts data of **Tables 1a & 2a** and partial cross sections data of **Tables 1b and 2b**, it is evident that the approximation to define the electron-electron exchange term introduced a large error to calculate the contribution of highly important p-wave (odd parity) e.g. the non-spherical orbitals.

In addition, a very large number of energy mesh-points in the energy interval 0.13 eV to 5.10 eV are used to calculate the s-wave elastic phase-shifts and the corresponding cross sections in search of Feshbach resonances using the exact-analysis of exchange, but no resonances are found in the system studying both the singlet and triplet partial phase shifts and corresponding cross sections.

To get the scattering lengths ($a^{\pm}$) for the s-wave elastic collision, the $k\cot\delta_0$ is plotted against $k^2$ in **Figure 4** for both the spin alignment S=0, singlet (+) and S=2, triplet (-) following the effective-range theory. In triplet channel the curve is almost straight line but in singlet channel it is slightly curved. It signifies that the contribution of the higher order terms (~$k^4$) in effective range theory is negligible in triplet (S=2) channel but it is not negligible in singlet (S=0) channel. In **Table 3** the computed scattering lengths and effective ranges are presented for both the singlet and triplet channels. They are again compared with available data [10,12-15]. Adhikari [12] and Chakraborty et al [13] used the similar type of coupled-channel approach. Adhikari [12] used a model potential to define the exchange effect and Chakraborty et al [13] used an approximation to describe the electron-electron exchange term. The present data with exact-exchange analysis are very close to the similar data reported in ref.[13]. It may appear that it is possible to approximate the most difficult electron-electron exchange term in Ps(1s)-Ps(1s) system by the positron-positron exchange term since both the electron and positron have equal masses and both the Ps atoms are in 1s-state which is spherically symmetric and the central potential V(**r**) is used, but the present detailed results made the idea that no approximate method is acceptable even for Ps(1s)-Ps(1s) elastic scattering. It is again mentioned in earlier presentation [29] that the use of non-spherical orbitals (e.g. p-states) in the coupled-channel method using an approximate form for the electron-electron exchange term would introduce large errors.

In **Figure 5** the integrated/ total elastic cross section $\sigma$ using the augmented-Born approximation and the quenching cross section $\sigma_q$ for the Ps(1s)-Ps(1s) elastic scattering are presented in the energy region 0 to 150 eV. A resonance like structure - a deep and a peak appears in the energy region 20-30 eV in both the total elastic and quenching cross sections in **Figure 5.** In **Figure 6** the ortho to para conversion ratios ($\sigma/\sigma_q$) are plotted against the incident energy and compared with the Ps-H system. It is evident from the figure that most of the annihilations occur at lower energies below 50 eV. All the data presented in **Figure 5 and 6** are completely new and no data are available to compare them.

In addition, the present code is not only useful to study Ps(1s)-Ps(1s) collision, but it is useful to study any two hydrogen-like atomic collisions. Using the present code and substituting the values of reduced mass and parameters of the wavefunctions of the Ps(1s)-H(1s) system, all the data of Ps(1s)-H(1s) elastic collision [20] were reproduced. The SEM is the simplest coupled-channel model that includes only the non-adiabatic short-range effects due to exchange but no long-range interaction. In the coupled-channel methodology [30] to include the effect of short-range forces atomic s-states are useful whereas to include the effect of long-range dipole-dipole van der Waals interaction atomic p-states are useful [21-23, 30]. So one needs to improve the present code including the excitation and continuum channels following the coupled channel methodology to get more reliable data at low and cold energies.

### 5. Conclusion

In conclusion an ab-initio static-exchange model (SEM) with a new computer code using FORTRAN programming language, is developed following the coupled channel methodology in momentum space to study Ps(1s)-Ps(1s) elastic collision considering the system as a four-body Coulomb problem in the CM frame system. Both the spin aligned: S=0, the singlet (+) and S=2, the triplet (-) are studied. All the Coulomb interaction terms in both the direct and exchange channels are calculated exactly. The present code is again useful to study any two hydrogen-like atomic collisions and the alkali atomic collisions at low and cold energies. The s-, p- and d- wave elastic phase shifts, the corresponding partial cross sections are being reported for the first time using an exact analysis of all the Coulomb interaction terms. The present data are compared with the available s-wave data using similar but approximate methods and are in close agreement. All p-wave (odd parity) phase shifts are $\pm \pi$ and corresponding cross sections are zero in the present exact-exchange analysis. The integrated/ total elastic cross section ($\sigma$), the quenching cross section ($\sigma_q$) and the conversion ratio ($\sigma/\sigma_q$) for the system are being reported for the first time. The data indicate the importance of quenching processes at lower energies below 50 eV. Again ortho to para conversion ratio is a useful quantity that could be measured. In addition a thorough search is made in search of Feshbach resonances but no resonances are found in the system. Both the scattering lengths are found positive. It is necessary to extend the code for more reliable data including the excitation and continuum channels with exact analysis of all the Coulomb interaction terms.

**Acknowledgement**

The author would feel happy to acknowledge the Department of Science & Technology (DST), India if financial support through Grant No. SR/WOSA/PS-13/2009 for the extended period.

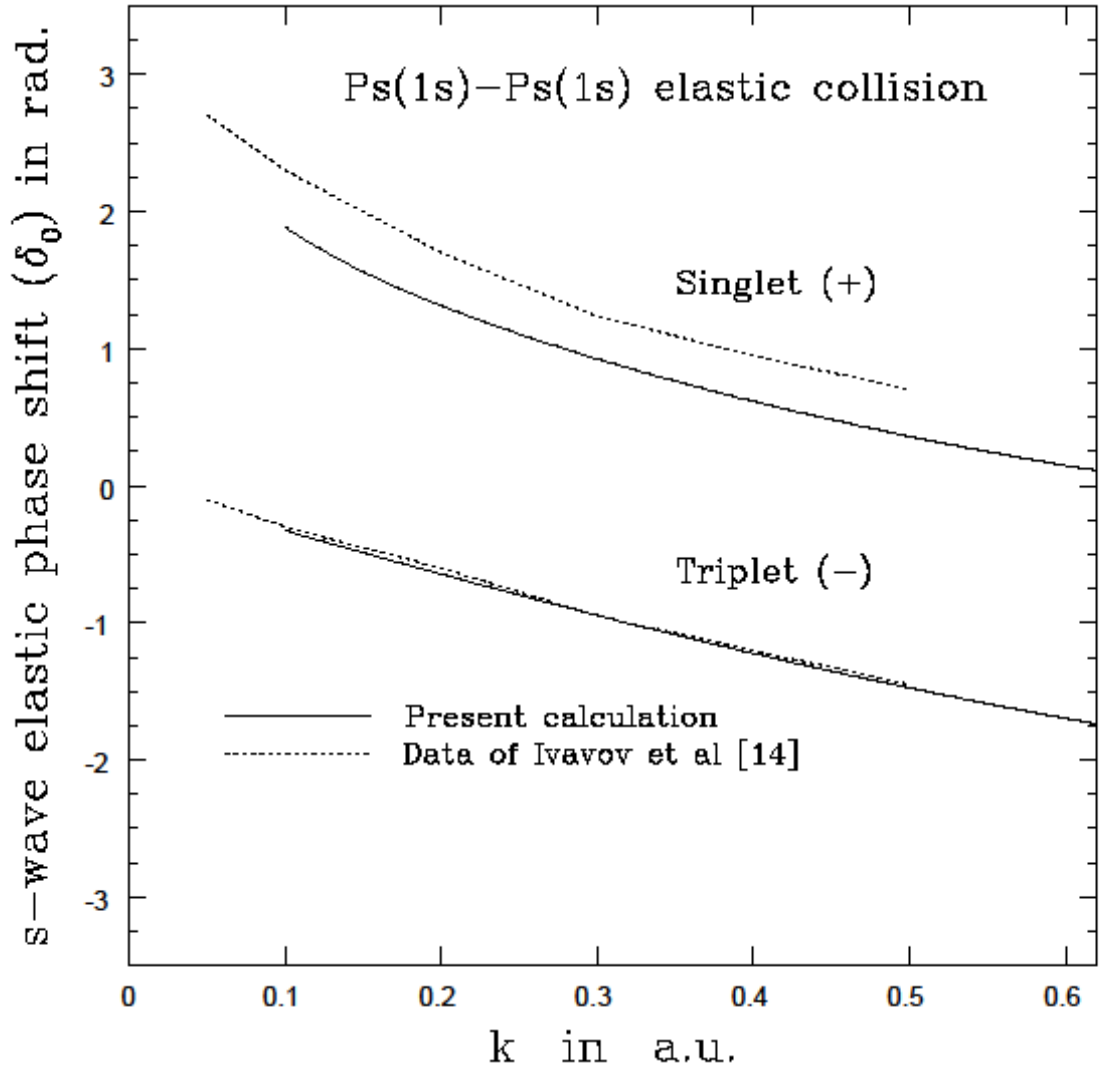

**Figure 2.** The s-wave elastic phase shifts in radian for both singlet (+) and triplet (-) channels in Ps(1s)-Ps(1s) scattering against the incident momentum k in a.u.

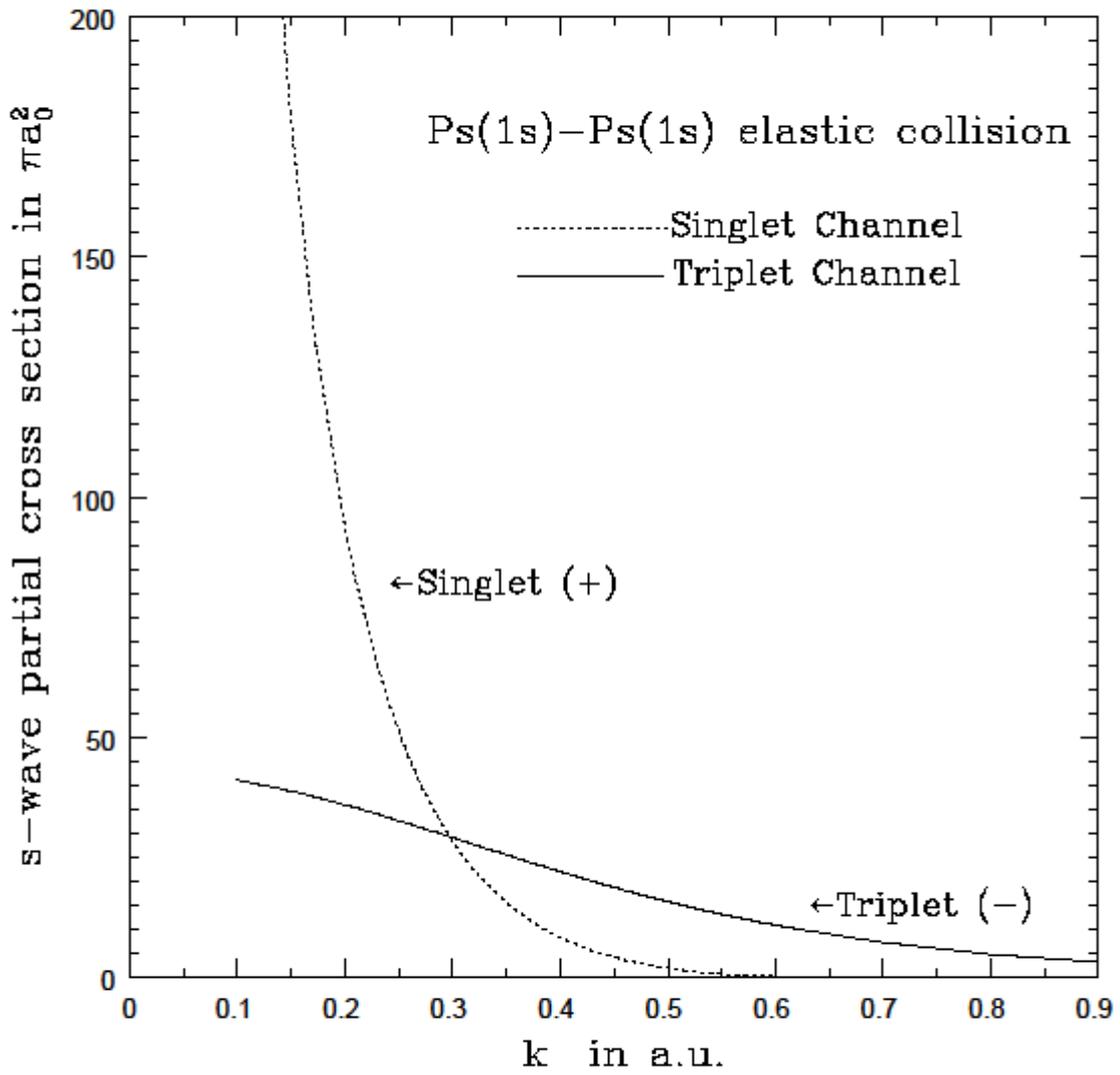

**Figure 3.** The s-wave elastic cross sections in $\pi a_0^2$ for both singlet (+) and triplet (-) channels in Ps(1s)-Ps(1s) scattering against the incident momentum k in a.u.

**Table 1a.** The s-, p- and d- wave phase shifts in radian for both singlet (+) and triplet (-) channels in Ps(1s) and Ps(1s) elastic scattering.

| k (a.u.) | Singlet (+) phase-shift in radian Ps(1s) and Ps(1s) system | | | Triplet (-) phase-shift in radian Ps(1s) and Ps(1s) system | | |
|---|---|---|---|---|---|---|
| | s-wave | p-wave | d-wave | s-wave | p-wave | d-wave |
| 0.1 | 1.887 | 3.142 | 3.141 | -0.325 | -3.142 | -3.141 |
| 0.2 | 1.313 | 3.142 | 3.141 | -0.643 | -3.142 | -3.141 |
| 0.3 | 0.926 | 3.142 | 3.138 | -0.944 | -3.142 | -3.137 |
| 0.4 | 0.616 | 3.142 | 3.129 | -1.222 | -3.142 | -3.127 |
| 0.5 | 0.359 | 3.142 | 3.114 | -1.474 | -3.142 | -3.109 |
| 0.6 | 0.147 | 3.142 | 3.095 | -1.697 | -3.142 | -3.085 |
| 0.7 | 3.118 | 3.142 | 3.073 | -1.892 | -3.142 | -3.059 |
| 0.8 | 2.986 | 3.142 | 3.052 | -2.059 | -3.142 | -3.035 |
| 0.9 | 2.890 | 3.142 | 3.035 | -2.200 | -3.142 | -3.017 |

**Table 1b**. The s-, p- and d- wave cross sections in $\pi a_0^2$ for both singlet (+) and triplet (-) channels in Ps(1s) and Ps(1s) elastic scattering.

| k (a.u.) | Singlet (+) cross-section in $\pi a_0^2$ Ps(1s) - Ps(1s) system | | | Triplet (-) cross-section in $\pi a_0^2$ Ps(1s) - Ps(1s) system | | |
|---|---|---|---|---|---|---|
| | s-wave | p-wave | d-wave | s-wave | p-wave | d-wave |
| 0.1 | 361.227 | 0.000 | 0.0000 | 41.016 | 0.000 | 0.0000 |
| 0.2 | 93.524 | 0.000 | 0.0002 | 35.966 | 0.000 | 0.0002 |
| 0.3 | 28.383 | 0.000 | 0.003 | 29.147 | 0.000 | 0.004 |
| 0.4 | 8.350 | 0.000 | 0.019 | 22.081 | 0.000 | 0.026 |
| 0.5 | 1.978 | 0.000 | 0.059 | 15.850 | 0.000 | 0.083 |
| 0.6 | 0.240 | 0.000 | 0.122 | 10.934 | 0.000 | 0.177 |
| 0.7 | 0.044 | 0.000 | 0.191 | 7.349 | 0.000 | 0.278 |
| 0.8 | 0.149 | 0.000 | 0.248 | 4.874 | 0.000 | 0.351 |
| 0.9 | 0.306 | 0.000 | 0.281 | 3.229 | 0.000 | 0.378 |

**Table 2a.** The s-, p- and d- wave phase shifts in radian for both singlet (+) and triplet (-) channels in Ps(1s) and Ps(1s) elastic scattering approximating the electron-electron interaction term.

| k (a.u.) | Singlet (+) phase-shift in radian Ps(1s) and Ps(1s) system | | | Triplet (-) phase-shift in radian Ps(1s) and Ps(1s) system | | |
|---|---|---|---|---|---|---|
| | s-wave | p-wave | d-wave | s-wave | p-wave | d-wave |
| 0.1 | 1.887 | 0.0153 | 3.141 | -0.326 | -0.0064 | -3.141 |
| 0.2 | 1.313 | 0.1210 | 3.141 | -0.643 | -0.0433 | -3.141 |
| 0.3 | 0.926 | 0.3564 | 3.138 | -0.944 | -0.1154 | -3.137 |
| 0.4 | 0.616 | 0.5967 | 3.129 | -1.222 | -0.2086 | -3.127 |
| 0.5 | 0.359 | 0.7125 | 3.114 | -1.474 | -0.3053 | -3.109 |
| 0.6 | 0.147 | 0.7245 | 3.095 | -1.697 | -0.3917 | -3.085 |
| 0.7 | 3.118 | 0.6843 | 3.073 | -1.892 | -0.4587 | -3.059 |
| 0.8 | 2.986 | 0.6233 | 3.052 | -2.059 | -0.5019 | -3.035 |
| 0.9 | 2.890 | 0.5568 | 3.035 | -2.200 | -0.5198 | -3.017 |

**Table 2b**. The s-, p- and d- wave cross sections in $\pi a_0^2$ for both singlet (+) and triplet (-) channels in Ps(1s) and Ps(1s) elastic scattering approximating the electron-electron interaction term.

| k (a.u.) | Singlet (+) cross-section in $\pi a_0^2$ Ps(1s) - Ps(1s) system | | | Triplet (-) cross-section in $\pi a_0^2$ Ps(1s) - Ps(1s) system | | |
|---|---|---|---|---|---|---|
| | s-wave | p-wave | d-wave | s-wave | p-wave | d-wave |
| 0.1 | 361.227 | 0.282 | 0.0000 | 41.016 | 0.0492 | 0.0000 |
| 0.2 | 93.524 | 4.368 | 0.0002 | 35.966 | 0.5620 | 0.0002 |
| 0.3 | 28.383 | 16.232 | 0.003 | 29.147 | 1.7692 | 0.004 |
| 0.4 | 8.350 | 23.685 | 0.019 | 22.081 | 3.2178 | 0.026 |
| 0.5 | 1.978 | 20.513 | 0.059 | 15.850 | 4.3372 | 0.083 |
| 0.6 | 0.240 | 14.641 | 0.122 | 10.934 | 4.8572 | 0.177 |
| 0.7 | 0.044 | 9.787 | 0.191 | 7.349 | 4.8024 | 0.278 |
| 0.8 | 0.149 | 6.389 | 0.248 | 4.874 | 4.3393 | 0.351 |
| 0.9 | 0.306 | 4.137 | 0.281 | 3.229 | 3.6558 | 0.378 |

**Table 3**. The computed scattering lengths and ranges in a.u. for both the singlet and triplet spin-states of system electrons in Ps(1s)-Ps(1s) elastic scattering.

| Ps(1s)-Ps(1s) collision | Singlet (+) | | Triplet (-) | |
|---|---|---|---|---|
| | Present data | Data of others | Present data | Data of others |
| Scattering length ( $a$ ) in a.u. → | 9.35 | 5.7 [10], 7.46 [12], 10.96[13], 8.44 [14], 8.26 [15], 9.15 [16] | 3.25 | 1.91 [10], 1.56 [12], 3.28[13], 3.00 [14], 3.02 [15], 3.0 [16] |
| Effective range ( $r_0$ ) in a.u. → | 7.36 | 7.27 [13] | 2.00 | 1.96 [13] |

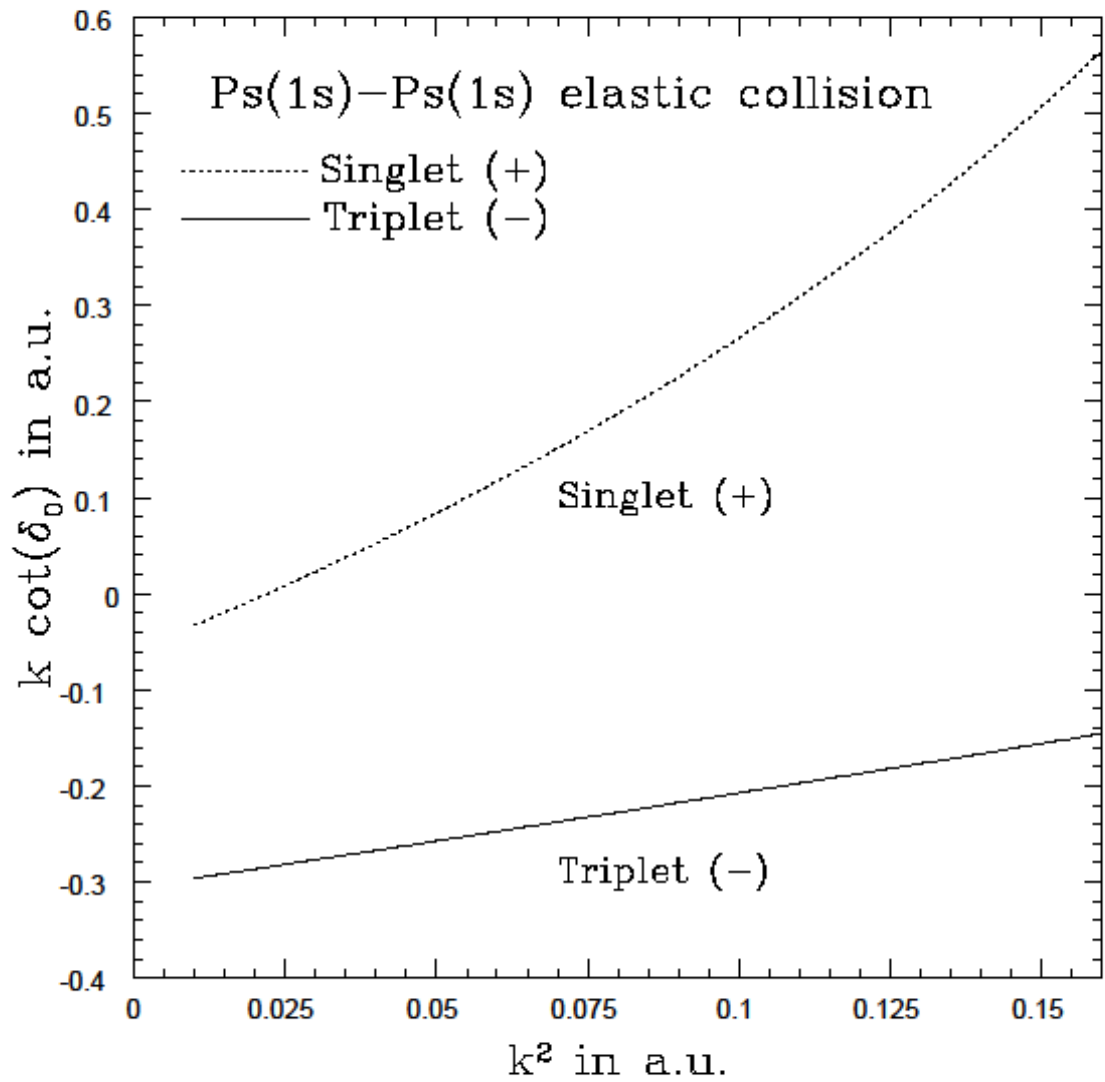

**Figure 4**. The plot of $k \cot \delta_0$ versus $k^2$ in a.u. of both the singlet (+) and triplet (-) spin-states of the two system electrons in Ps(1s)-Ps(1s) elastic scattering.

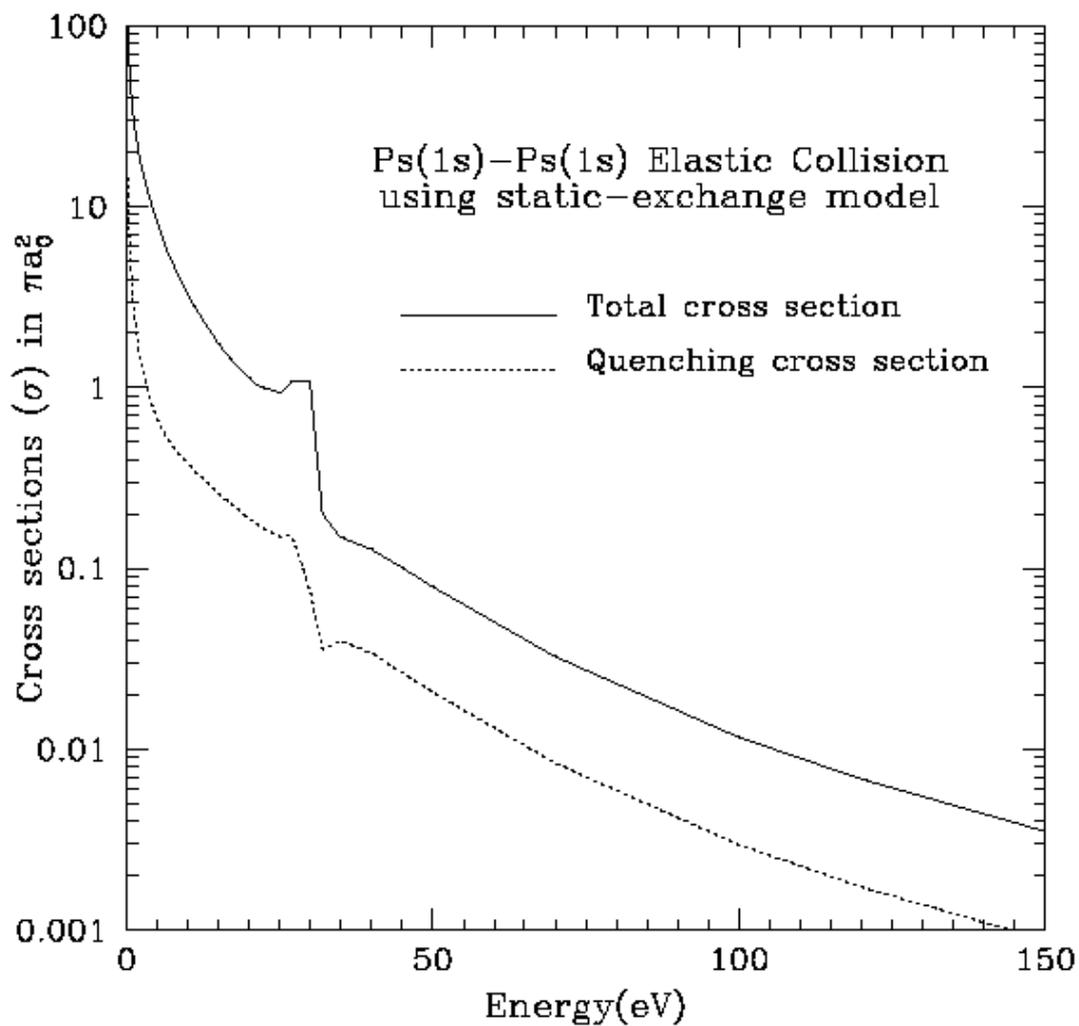

**Figure 5.** The variation of integrated /total elastic cross sections and the quenching cross section in $\pi a_0^2$ of Ps(1s)-Ps(1s) scattering with energy in eV.

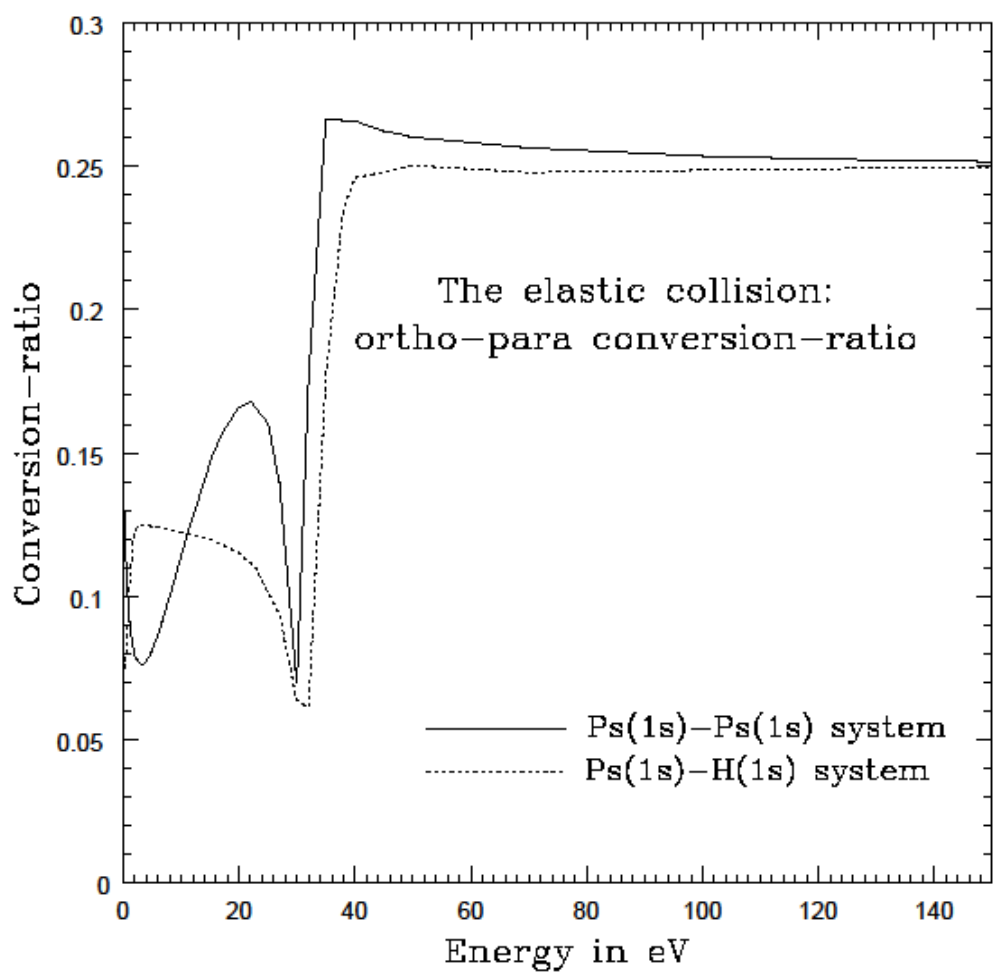

**Figure 6.** The comparison of the variation of ortho to para conversion ratios ($\sigma_q/\sigma$) of Ps(1s)-Ps(1s) and Ps(1s)-H(1s) systems with energy.